\documentclass[12pt,journal]{IEEEtran}
\usepackage{epsfig}
\usepackage{amssymb}
\usepackage{graphicx}
\usepackage{graphics}
\usepackage{amsmath}
\usepackage{setspace}
\onecolumn
\title{
Information Capacity of %Gaussian Channel with
   an Energy Harvesting Sensor Node}

\author{R Rajesh,\thanks{Preliminary versions of parts of this paper appear  in ISIT 2011, Asilomer 2011 and  Globecom 2011. 
 R Rajesh is  with Center for Airborne Systems, DRDO, Bangalore. Vinod Sharma is with ECE dept. Indian Institute of Science, Bangalore. Pramod Viswanath is with  Electrical and Computer Engineering dept. at University of Illinois, Urbana-Champaign. This work was done when Prof. Viswanath was visiting Indian Institute of Science.  Email:rajesh81r@gmail.com, vinod@ece.iisc.ernet.in, pramodv@illinois.edu.}
 \thanks{This work is partially supported by a grant from ANRC to Prof. Sharma.}
\thanks{The visit of Prof. Viswanath was supported by DRDO-IISc Programme on Advanced Research in Mathematical Engineering.}
\and Vinod Sharma % \IEEEmembership{Senior Member IEEE} 
and \and Pramod Viswanath 
}

\begin{document}
\maketitle
\thispagestyle{empty}
\pagestyle{empty}
\begin{abstract}
Energy harvesting sensor nodes are gaining popularity due to their ability to improve the network life time and  are becoming a preferred choice supporting ``green communication". In this paper we focus on communicating reliably over an AWGN channel using such an energy harvesting sensor node. An important part of this work involves appropriate modeling of the energy harvesting, as done via  various practical architectures. Our main result is the characterization of the Shannon capacity of the communication system. The key technical challenge involves dealing with the dynamic (and stochastic) nature of the (quadratic) cost of the input to the channel. As a corollary, we find close connections between the capacity achieving energy management policies and the queueing theoretic throughput optimal policies. 
%We  also obtain the capacity when energy conserving  sleep-wake modes are supported and an achievable rate for the system with  inefficiencies in energy  storage. Next we extend these results to a fading AWGN channel with perfect/no channel state information at the transmitter.  We also combine the information theoretic and queueing-theoretic models for the above scenarios. Finally, we provide achievable rates when the nodes have finite buffer to store the harvested energy.
\end{abstract}

\section{Introduction}
\label{intro}

Sensor nodes are often  deployed for monitoring a random field. These nodes are characterized by  limited battery power, computational resources  and storage space. Once deployed, the battery of these nodes are often not changed because of the inaccessibility of these nodes. 
 Nodes could possibly use larger batteries but with increased weight, volume and cost.  Hence when the battery of a node is exhausted, it is not replaced and the node dies. When sufficient number of nodes die, the network may not be able to perform its designated task. Thus the life time of a network is an important characteristic of a sensor network (\cite{bhardwaj}) and it depends on the life time of a node.

The network life time can be improved by  reducing the energy intensive tasks, e.g., reducing the number of bits to transmit (\cite{pradhan}, \cite{baek}), making a node to go into power saving modes (sleep/listen) periodically (\cite{sinha}), using energy efficient routing (\cite{woo}, \cite{ratnaraj}), adaptive sensing rates and multiple access channel  (\cite{ye}). Network life time can also be increased by suitable architectural choices like the tiered system (\cite{tire}) and redundant placement of nodes (\cite{red}).

Recently new techniques of  increasing network life time by increasing the life time of  the battery is gaining popularity. This is made possible by energy harvesting techniques (\cite{kansal}, \cite{niyato}). Energy harvestesr harness energy from the environment or other energy sources ( e.g., body heat) and convert them to electrical energy. Common energy harvesting devices are solar cells, wind turbines and  piezo-electric cells, which extract energy from the environment. Among these,  harvesting solar energy through photo-voltaic effect seems to have emerged as a technology of choice for many sensor nodes (\cite{niyato}, \cite{raghunathan}). Unlike for a battery operated sensor node, now there is potentially an \textit{infinite} amount of energy available to the node. However, the source of energy and the energy harvesting device may be such that the energy cannot be generated at all times (e.g., a solar cell).  Furthermore the rate of generation of energy can be limited. Thus one may want to match the energy generation profile of the harvesting source with the energy consumption profile of the sensor node. If the energy can be \textit{stored} in the sensor node then this matching can be considerably simplified. But the energy storage device may have limited capacity. The energy consumption policy  should be designed in such a way that the node can perform satisfactorily for a long time, i.e., energy starvation at least, should  not be the reason for the node to die. In \cite{kansal} such an energy/power management scheme is called  \textit{energy neutral operation}.

%We study the Shannon capacity of such an energy harvesting sensor node transmitting over an Additive White Gaussian Noise (AWGN) Channel. We provide the capacity under various energy buffer conditions and show that the capacity achieving policies are related to throughput optimal policies(\cite{vinod1}). We also study generalizations of this system with  various inefficiencies in storage and different modes of operation. We also generalize these results to fading channels.

In the following we survey the relevant literature. Early papers on energy harvesting in sensor networks are \cite{kansal1} and \cite{rahimi}. A practical solar energy harvesting sensor node prototype is described in \cite{jiang}. In  \cite{kansal} various deterministic  models for energy generation and energy consumption profiles are studied and provides conditions for energy neutral operation. In \cite{jaggi}  a sensor node is considered which is sensing certain interesting events. The authors study optimal sleep-wake cycles such that event detection probability is maximized. A recent survey on energy harvesting is  \cite{erkal}.

Energy harvesting can be  often divided into two major architectures (\cite{jiang}). In {\it{Harvest-use}}(HU), the harvesting system directly powers the sensor node and when sufficient energy is not available the node is disabled. In {\it{Harvest-Store-Use}} (HSU) there is a storage device that stores the harvested energy and also powers the sensor node. The storage can be single or double staged (\cite{kansal},~\cite{jiang}).

Various throughput and delay optimal energy management policies for energy harvesting sensor nodes are provided in \cite{vinod1}.  The energy management policies in \cite{vinod1} are extended in  various directions in \cite{vinod2} and \cite{vinod3}. For example, \cite{vinod2} also provides some efficient MAC policies for energy harvesting nodes. In \cite{vinod3} optimal sleep-wake policies are obtained for such nodes. Furthermore, \cite{vinod4} considers jointly optimal routing, scheduling and power control policies for networks of energy harvesting nodes. Energy management policies for finite data and energy buffer are provided in \cite{kok}. Reference \cite{elsajsc} provides optimal energy management policies and energy allocation over source acquisition/compression and transmission. 

%Various attempts to unify information and queuing theory for multi-access systems are also given in \cite{gal} and \cite{effros}. The equivalence of  information theoretic capacity region and stability region of a multiple access channel is shown in \cite{effros}.

In a recent contribution, optimal energy allocation policies over a finite horizon and  fading channels are studied  in \cite{sg}.  Relevant literature for models combining information theory and queuing theory are \cite{effros} and \cite{gal}.

The capacity of a fading Gaussian channel with channel state information (CSI) at the transmitter and receiver and at the receiver alone are provided in \cite{varia}. It was shown that optimal power adaptation when CSI is available both at the transmitter and the receiver is `water filling' in time. 
%An excellent survey on fading channels is provided in  \cite{proakis}.  

%%%%%%%%%%%%%%%%%%%%%%%%%%%%%%%%%%%%%%%%%%%%%%
% Changes made by Pramod
%%%%%%%%%%%%%%%%%%%%%%%%%%%%%%%%%%%%%%%%%%%%%%%%
Information-theoretic capacity of an energy harvesting system has been considered previously in \cite{uluk} and \cite{ncc11} independently. It was shown that the capacity of the energy harvesting AWGN channel with an unlimited battery is equal to the capacity with an average power constraint equal to average recharge rate.  In \cite{ncc11} the proof technique used is based on AMS sequences \cite{gray} which is different from that used in \cite{uluk}. 

%This result is presented in this paper (as Theorem 1) to highlight two features: a variant of an achievable scheme from \cite{uluk} that is also capacity achieving (\cite{uluk} presents two achievable schemes Save-and-Transmit and Best-Effort-Transmit; the achievability we use in this paper is a slight variation of the Best-Effort-Transmit scheme in \cite{uluk}),  as well as introducing  a different technique to prove the achievability of the proposed rates; in particular, we use AMS sequences \cite{gray}. Our converse proof, conducted independently,  is the same as in \cite{uluk}. 

Our main contributions are in considering significant extensions to the basic energy harvesting system by considering processor energy, energy inefficiencies (and finally channel fading). 
We  compute the capacity when the energy is  consumed in other activities at the node (e.g., processing, sensing, etc) than transmission. This issue of energy consumed in processing in the context of the usual AWGN channel (i.e., without energy harvesters) is addressed in \cite{meda}. Finally we provide the achievable rates when there are storage inefficiencies. We show that the throughput optimal policies provided in \cite{vinod1} are related to the  capacity achieving policies provided here. We also extend the results to a scenario with fast fading. Further we combine the information theoretic and queueing-theoretic models for the above scenarios. Finally, we provide achievable rates when the nodes have finite buffer to store the harvested energy. Our results can be useful in the context of green communication (\cite{green1},~\cite{green2}) when solar and/or wind energy can be used by a base station (\cite{wind}).%After submission of the first version to {\tt arxiv}, independently \cite{uluk} appeared on {\tt{arxiv}} which addresses this problem for a special case of the model considered in Theorem 1 of this paper. Their method of proof is entirely different. Our results can be useful in the context of green communication (\cite{green1},~\cite{green2}) when solar and/or wind energy can be used by a base station (\cite{wind}).

%Initially we  assume that most of the energy is consumed in transmission only. We will relax this assumption later on. We find the capacity of such a system. We also relate the achievable region with the conditions required for energy neutral operation of the system, i.e., when the system can work forever and the data queue is stable (if data queue is available).
System level power consumption in wireless systems including energy expended in  decoding is provided in \cite{sah}. Related literature for conserving energy but without energy harvester is \cite{p1}-\cite{p2}. In \cite{p1} an explicit model for power consumption at an idealized decoder is studied. Optimal constellation size for uncoded transmission subject to peak power constraint is given in  \cite{p4}. Reference \cite{p2} characterizes the capacity when the transmitter and the receiver probe the state of the channel. The probing action is cost constrained. 

The paper is organized as follows. Section \ref{model} describes the system model. Section \ref{stability} provides the capacity for a single node under idealistic assumptions. Section \ref{simulation} takes into account the energy spent on sensing, computation etc. and proposes capacity achieving  sleep-wake schemes. Section \ref{opt} obtains efficient policies with inefficiencies in the energy storage system. Section \ref{model1} studies the capacity of the energy harvesting system transmitting over a fading AWGN channel. Section VII combines the information-theoretic and queueing-theoretic formulations. Section VIII provides achievable rates for the practically interesting case of finite buffer. Section \ref{conclude} concludes the paper.

\section{Model and notation}
\label{model} 
\begin{figure}[h]
\begin{center}
\includegraphics[height= 1.6  in, width=4 in]{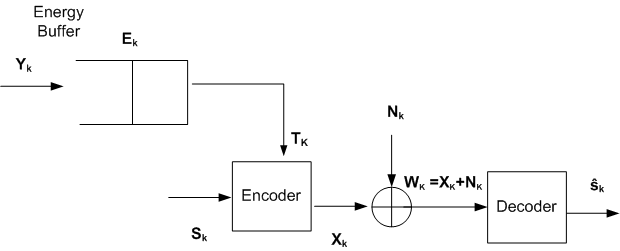}
\end{center}
\begin{center}
\caption{The model} \label{fig1}
\end{center}
\end{figure}
In this section we present our model for a single energy harvesting sensor node.We consider a sensor node (Fig. \ref{fig1}) which is sensing and generating data to be transmitted to a central node via a discrete time AWGN channel.  We assume that transmission consumes most of the energy in a sensor node and ignore other causes of energy consumption (this is true for many low quality, low rate sensor nodes (\cite{raghunathan})). This assumption will be removed in Section \ref{simulation}. The sensor node is able to replenish energy by $Y_k$  at time $k$. The energy available in the node at time $k$ is $E_k$. This energy is stored in an energy buffer with an infinite capacity. In this section the fading effects are not considered; however this issue is addressed in Section \ref{model1}. %We will consider the effect of finite capacity later.

The node uses energy $T_k$ at time $k$ which depends on $E_k$ and $T_k \le E_k$ . The process $ \{ E_k \}$ satisfies
\begin{eqnarray}
E_{k+1}  = (E_k - T_k) + Y_k. \label{eqn2}
\end{eqnarray}

We will assume that  $\{ Y_k \}$  is  stationary ergodic. This assumption is general enough to cover most of the stochastic models developed for energy harvesting.  Often the energy harvesting process will be time varying (e.g., solar cell energy harvesting will depend on the time of day). Such a process can  be approximated by piecewise stationary processes. As in \cite {vinod1}, we can indeed consider $\{Y_k\}$ to be periodic, stationary ergodic.

The encoder receives a message $S$ from the node and generates an $n$-length codeword to be transmitted on the AWGN channel. The channel output  $W_k=X_k+N_k$ where $X_k$ is the channel input at time $k$ and $N_k$ is independent, identically distributed (\emph{iid}) Gaussian noise with zero mean and variance $\sigma^2$ (we denote the corresponding Gaussian density by $\mathcal{N}(0,\sigma^2))$. The decoder receives $W^n \stackrel{\Delta}{=}(W_1,...,W_n)$ and reconstructs  $S$ such that the probability of decoding error is minimized.

We will obtain the information-theoretic capacity of this channel. This of course assumes that there is always data to be sent at the sensor node (this assumption will be removed in section VII). This channel is essentially different from the usually studied systems in the sense that the transmit power and coding scheme can depend on the energy available in the energy buffer at that time.

A possible generalization of our model is that the energy $E_k$ changes at a slower time scale than a channel symbol transmission time, i.e., in equation \eqref{eqn2} $k$ represents a time slot which consists of $m$ channel uses, $m \ge 1$. We comment on this generalization  in Section III (see also Section VII).

%%%%%%%%%%%%%%%%%%%%%%%%%%%%%%%%%%%%

%%\emph{Asymptotic Mean Stationary} (\cite{gray}): A dynamical system $(\Omega, \mathcal{B}, P, T)$ is said to be stationary if $
%P(T^{-1}G)= P(G)$ for all $ G \in \mathcal{B}$. It is said to be
%\emph{asymptotically mean stationary} (AMS) if the limit $
%\overline{P}(G)= \lim_{n \to \infty} \frac{1}{n} \sum_{k=0}^{n-1}
%P(T^{-1}G)$ exists for all $ G \in \mathcal{B}$. $\overline{P}$ is
%called the stationary mean.

%%\emph{Pinskar Information Rate} (\cite{gray}): Given an AMS pair
%random process $\{X_n; Y_n\}$ with standard alphabets $A_X$ and
%$A_Y$, the pinskar information rate is defined as $I^{*} (X;Y)=
%sup_{q,r} \overline{I} (q(X); r(Y))$
% where the supremum is over all quantizes $q$ of
%$A_X$  and $r$ of $A_Y$ and $\overline{I}(q(X);r(W)) = lim~sup_{n \to \infty}  I(q(X^n);r(W^n))$. Also, $ I^*(X;Y) < \overline{I}(X;Y)$. The two are equal if the alphabets are finite.

%We also need the following Lemma for proving the achievability part.
%%%%%%%%%%%%%%%%%%%%%%%%%%%%%%

\section{Capacity for the Ideal System}
 \label{stability}
In this section we obtain the capacity of the channel with an energy harvesting node under ideal conditions of  infinite energy buffer and energy consumption in transmission only.

The system starts at time $k=0$ with an empty energy buffer and $E_k$ evolves with time depending on $Y_k$ and $T_k$. Thus $\{E_k,~k \ge 0\}$ is not stationary and hence $\{T_k\}$ may also not be stationary. In this setup, a reasonable general assumption is to expect $\{T_k\}$ to be asymptotically stationary. Indeed we will see that it will be sufficient for our purposes. These sequences are a subset of Asymptotically Mean Stationary (AMS) sequences , i.e., sequences $\{T_k\}$ such that 
\begin{equation}
\lim_{n \to \infty} \frac{1}{n} \sum_{k=1}^n P[T_k \in A]= \overline{P}(A)
\end{equation}
exists for all measurable $A$. In that case $\overline{P}$ is also a probability measure and is called the \emph{stationary mean} of the AMS sequence (\cite{gray}).

If the input $\{X_k\}$ is AMS and ergodic, then it can be easily shown that for the AWGN channel $\{(X_k,W_k),~k \ge 0\}$ is also AMS an ergodic. In the following theorem  we will show that the channel capacity of our system is (\cite{gray})
\begin{equation}
\label {eqnn1}
C= \sup_{p_x} \overline{I}(X;W)= \sup_{p_x} ~ \limsup_{n \to \infty}\frac{1}{n} I(X^n,W^n)
\end{equation}
where $\{X_n\}$ is an AMS sequence, $X^n=(X_1, ..., X_n)$ and the supremum is over all possible AMS sequences $\{X_n\}$. In other words,  one can find a sequence of codeword�s with code length $n$ and rate $R$ such that  the average probability of error goes to zero as $n \to \infty$ if and only of $R<C$.

% We need the following definition. Although for $C$ we need mutual information rate $\overline{I}(X;W)$, $I^{*}(X;W)$ defined below is easier to handle.

%\emph{Pinsker Information Rate} (\cite{gray}): Given an AMS  random process $\{(X_n, W_n)\}$ with
 %standard alphabets (Borel subsets of Polish spaces) $A_X$ and $A_W$, the Pinsker information rate is defined as
% $I^{*} (X;W)= \sup_{q,r} \overline{I} (q(X); r(W))$ where the supremum is over all quantizers $q$ of
%$A_X$  and $r$ of $A_W$ and $\overline{I}(q(X);r(W)) = \limsup_{n \to \infty}  I(q(X^n);r(W^n))/n$.

%It is known that, $ I^*(X;W) \le \overline{I}(X;W)$. The two are equal if the alphabets are finite.

%We also need the following Lemma for proving the achievability of the capacity of the channel. This Lemma holds for $I^*$ but not for $\overline{I}$.

%{\bf Lemma 1} (\cite{gray}, Lemma  6.2.2):  Let  $\{(X_n, W_n)\}$  be  AMS  with distribution $P$ and stationary mean $\overline{P}$. Then $I^*_P(X;W) = I^*_{\overline{P}}(X;W)$.

{\bf Theorem 1}: For the energy harvesting system, the capacity $C = 0.5~\log(1+\frac{E[Y]}{\sigma^2}) $.

{\bf Proof:}  See Appendix A. This result has also appeared in \cite{uluk}. The achievability proofs are somewhat different (both the scheme itself as well as the technical approach to the proof). 
\vspace{0.2cm}

Thus we see that the capacity of this channel is the same as that of a node with average energy constraint $E[Y]$, i.e., the hard energy constraint of $E_k$ at time $k$ does not affect its capacity.
The capacity achieving signaling in the above theorem is truncated $iid$ Gaussian
 with zero mean and variance $E[Y]$ where the truncation occurs due to the energy limitation $E_k$ at time $k$.
 The same capacity is obtained for any other initial energy $E_0$ (because then also our signaling scheme leads to an AMS sequence with the same stationary  mean).

%%%%%%%%%%%%%%%%%%%%%%%%%%%%%%%%%%%%%%%%%%%%%%
% Changes made by Pramod
%%%%%%%%%%%%%%%%%%%%%%%%%%%%%%%%%%%%%%%%%%%%%%%%
The scenario when there is no energy buffer to store the harvested energy (Harvest-Use) was studied extensively in \cite{uluk2}, which calculated the capacity to be 
 $C=\max_{p_x}I(X;W)\le 0.5~ \log (1+ E[Y]/\sigma^2)$. We mention this result  in some detail (and variations) since this material will be used in developing later sections. The last inequality is strict unless $X_k$ is $\mathcal{N}(0,E[Y])$ and $Y_k$ is also known at the receiver at time $k$.  Then $X^2=Y$ and hence $Y_k$ is chi-square distributed with degree 1. If $Y_k\equiv E[Y]$ then the capacity will be that of an AWGN channel with peak and average power constraint $=E[Y]$. This problem is addressed  in \cite{peak1}, \cite{peak2}, \cite{peak3} and the capacity achieving distribution is finite and discrete. Let  $X(y)$ denote a random variable  having distribution that  achieves capacity with peak power $y$. Then, for the case when information about $Y_k$ is also available at the decoder at time $k$, the capacity of the channel when    $\{Y_k\}_{k \geq 1}$ is $iid$ is

\begin{equation}
C= E_Y[I(X(Y);W)].\label{fin}
\end{equation}
For small $y$, $X^2(y)=y$. This result can be extended to the case when $\{Y_k\}$ is stationary ergodic. Then the right side of \eqref{fin} will be replaced by the information rate of $\{X_k(Y_k),W_k\}$.
In conclusion,  having some energy buffer to store the harvested energy almost always strictly increases the capacity of the system (under ideal conditions of this section).

Next we extend this result to the case when only partial information about  $Y_k$ is available at the encoder and the decoder at time $k$ (causally). The interesting case of  $Y_k$ information being perfectly available at the encoder and not at the decoder is a special case of  this set up. The channel is given in Fig. \ref{fig_new} where $V^{(t)}_k$ and  $V^{(r)}_k$ denote the partial information about $Y_k$ at the encoder and the decoder respectively. For simplicity, we will assume $\{Y_k\}$ to be $iid$. The capacity of this channel can be obtained from the capacity of a  state dependent channel with partial state information at the encoder and the decoder (\cite{now}):  

\begin{figure}[h]
\begin{center}
\includegraphics[height= 1.5  in, width=4 in]{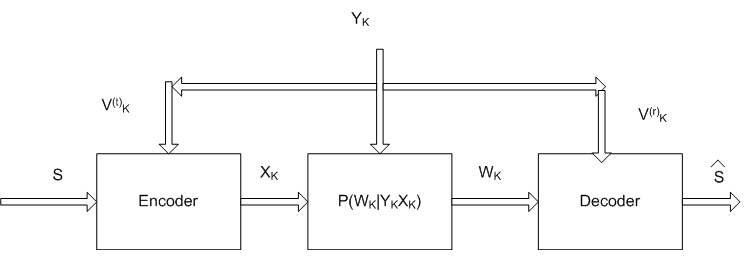}
\caption{Capacity with no buffer and partial  energy harvesting information} \label{fig_new}
\end{center}
\end{figure}

\begin{equation}
C= \sup_{P_T(.)} I(T;W|V^{(r)}), \label{fin123}
\end{equation}
where the supremum is over distributions $P_T(.)$ of continuous functions, $T \in \mathcal{T}$,  $T: \mathcal{V}^{(t)} \to \mathcal{X}$ where $\mathcal{V}^{(t)}$ and  $\mathcal{X}$ denote the  sets in which $Y$ and $X$ take values. $\mathcal{T}$ denotes the set of all $|\mathcal{X}|^{|\mathcal{V}^{(t)}|}$ functions from $\mathcal{V}^{(t)}$ to $\mathcal{X}$. Also, $T$ is independent of $V^{(t)}$ and $V^{(r)}$.  The capacity when $Y_k$ is not available at the decoder but perfectly known at the encoder is obtained by substituting $V^{(t)}_k=Y_k$ and $V^{(r)}_k=\phi$ in \eqref{fin123}.

In \cite{vinod1}, a system with  a data buffer at the node which
stores data sensed by the node before transmitting it, is considered.
 The stability region (for the data buffer) for the 'no-buffer' and 'infinite-buffer'
corresponding to the harvest-use and harvest-store-use architectures are provided.
The throughput optimal policies in \cite{vinod1} are  $T_n= \min(E_n;E[Y]-\epsilon)$ for
 the  infinite energy buffer and  $T_n = Y_n$ when there is no energy buffer.
  Hence  we see that the Shannon capacity achieving energy management policies provided here are
 close to the throughput optimal policies in \cite{vinod1}.  Also the capacity is the
 same as the maximum throughput obtained in the data-buffer case in \cite{vinod1} for the infinite buffer architecture. In section VII we will connect further this model with our information theoretic model studied above.

 Above we considered the cases when there is infinite
 energy buffer or when there is no buffer at all.
However, in practice often there is a finite energy buffer to store. This case is considered in Section VIII and we provide achievable rates.

Next we comment on the capacity results when \eqref{eqn2} represents $E_{k+1}$ at the end of the $k$th slot where a slot represents $m$ channel uses. In this case energy $E_k$ is available not  for one channel use but for $m$ channel uses. This relaxes our energy constraints. Thus if $E[Y]$ still denotes mean energy harvested per channel use, then for infinite buffer case the capacity remains same as in Theorem 1.% However the no buffer case becomes like a finite buffer case studied above.

\section{Capacity with Processor Energy (PE)}
\label{simulation}
Till now we have assumed that all the energy that a node consumes is for transmission.
However, sensing, processing and receiving (from other nodes) also require significant energy,
 especially in recent higher-end sensor nodes (\cite{raghunathan}).
We will now include the energy consumed by sensing and processing only.% We will see that if a node has an energy saving mode, the achievable rate can be improved. For simplicity, we will initially assume that  the node is always in one energy mode (i.e., there is no energy saving mode).

We  assume that energy $Z_k$  is consumed by the node (if $E_k \ge Z_k$) for
 sensing and processing  at time instant $k$. Thus, for transmission at time $k$, only $E_k-Z_k$ is available.
 $ \{ Z_k \}$ is assumed a stationary, ergodic sequence. The rest of the system is as in Section II.

First we extend the achievable policy in Section \ref{stability} to
incorporate this case. The signaling scheme $X_k=sgn(X_k')\min(\sqrt{E_k},|X_k'|)$ where $\{X_k'\}$ is $iid$
Gaussian with zero mean and variance $E[Y]-E[Z]-\epsilon$  achieves
the rate
\begin{equation}
R_{PE}= 0.5~\log\left(1+\frac{E[Y]-E[Z]-\epsilon}{\sigma^2}\right).\label{abc}
\end{equation}

If the sensor node has two modes: Sleep and Awake then the achievable rates can be improved. The sleep mode is a power saving
mode in which  the sensor only harvests energy and performs no other functions so that the energy consumption is minimal (which will be ignored). If $E_k < Z_k$ then we assume that the node will sleep at time $k$. But to optimize its transmission rate it can sleep at other times also. We consider a policy called \emph{randomized sleep policy} in \cite{vinod3}. In  this policy at each time instant $k$ with $E_k \ge Z_k$ the sensor chooses to sleep with probability $p$ independent of all other random variables.  We will see that such a policy can be capacity achieving in the present context. %We show that this policy is capacity achieving in the class of randomized sleep-wake rules studied in \cite{vinod3}.

With the sleep option we will show that the capacity of this system is
\begin{equation}
C= \sup_{p_x: E[b(X)] \le E[Y]} I(X;W), \label{impe}
\end{equation}
where $b(x)$ is the cost of transmitting $x$ and equals
\begin{eqnarray*}
b(x)= \begin{cases}
x^2+\alpha,~& \text{if $|x| >0$},\\
0,~& \text{if $|x| = 0$},
\end{cases}
\end{eqnarray*}
and $\alpha= E[Z]$. Observe that if we follow a policy that unless the node transmits, it sleeps,
 then $b$ is the cost function.  An optimal policy will have this characteristic.
  Denoting the expression in \eqref{impe} as $C(E[Y])$, we can easily check that $C(.)$ is a non-decreasing
  function of $E[Y]$. We  also show below that $C(.)$ is concave. These  facts will be used in proving that \eqref{impe} is the capacity of the system.

To show concavity, for $s_1, s_2 >0$ and $0 \le \lambda \le 1$ we want to show that $C( \lambda s_1 + (1-\lambda) s_2) \ge \lambda C(s_1)+(1-\lambda) C(s_2)$. For $s_i$, let $C_i$ be the capacity achieving codebook, $i=1,2$. Use $\lambda$  fraction of time $C_1$ and $1-\lambda$ fraction $C_2$. Then the rate achieved is $\lambda C(s_1)+(1-\lambda) C(s_2)$ while the average energy used is $\lambda s_1+(1-\lambda) s_2$. Thus, we obtain the inequality showing concavity.

{\bf Theorem 2} For the energy harvesting system with processing energy,

\begin{equation}
C= \sup_{p_x: E[b(X)] \le E[Y]} I(X;W) \label{impee}
\end{equation}
is the capacity for the system.

{\bf Proof:}  :  See Appendix B.

\vspace{0.2cm}

%In the class of policies where $\overline{I}=I^*$ we can show that $R$ is the capacity. Some conditions when $\overline{I}=I^*$ are given in \cite{gray}.

It is interesting to compute the capacity \eqref{impee} and the capacity achieving distribution. Without loss of generality, the node sleeps with probability $p, (0 \leq p \leq 1)$ and with probability $(1-p)$ the node transmits with a distribution $F_t(.)$.  We can write the overall input distribution, $F_{in}(.)$, as a mixture distribution
\begin{equation*}
F_{in}(.)=pu(.)+(1-p)F_t(.),
\end{equation*}
where $u(.)$ denotes the unit step function. The corresponding output density function $f_W(.;F_t)=pf_N(.)+(1-p)\int f_N(.-s)dF_t(s)$, is the convolution of $F_{in}(.)$ and $f_N(.)$ where  $f_N(.)$ is $\mathcal{N}(0,\sigma^2)$. The mutual information $I(X;W)$ in \eqref{impee} can be written as 
\begin{equation*}
I(F_t) \triangleq I (X;W) =ph(0;F_t)+(1-p)\int h(x;F_t)dF_t(x)-h(N),
\end{equation*}
where $h(N)$ is the differential entropy of noise $N$ and  $h(x;F_t)$ is the marginal entropy function defined as 
\begin{equation*}
h(x;F_t) =-\int f_N(w-x)\log (f_W(w;F_t))dw.
\end{equation*}

Capacity computation can be formulated as a constrained maximization problem,
\begin{equation}
\label{omega}
\sup_{F_t \in \Omega}I(F_t),
\end{equation}
where $\Omega \triangleq \{F_t: F_t \text{ is a cdf and} \int s^2dF_t(s)\leq \beta_p\}$ and $\beta_p \triangleq \frac{E[Y]}{(1-p)}-\alpha$. $\Omega$  is the space of all distribution functions with finite second moments and is endowed with the topology of weak$^*$ convergence. This topology is metrizable with Prohorov metric (\cite{bass}). It is easy to see that $\Omega$ is a compact, convex topological space. The compactness of $\Omega$ is a consequence of the second moment constraint of the distribution function which makes it tight and Helly's theorem. The objective function $I(F_t)$ is a strictly concave map  from $\Omega$ to $\mathbb{R}^+$, the positive real line. We can show that $I(F_t)$ is a continuous function in the weak$^*$ topology and $I(F_t)$ admits a weak derivative \cite{peak1}. Then there is a unique distribution $F_{t0}$ that optimizes \eqref{omega}. The weak derivative of $I(F_t)$ with respect to $F_t$ at the optimum distribution $F_{t0}$ is
\begin{equation*}
I'_{F_{t0}}(F_t)=ph(0;F_{t0})+(1-p) \int h(x;F_{t0})dF_t(x)-h(N)-I(F_{to}).
\end{equation*}
Here, $I(F_{to})$ is the capacity of the channel. Using KKT conditions we get sufficient and necessary conditions as  $I'_{F_{t0}}(F_t) \leq 0$ and the conditions can be simplified using the techniques in \cite{peak1}, \cite{abu} as 

\begin{flalign}
\label{kkt1}
Q(x) \triangleq  (1-p) h(x;F_{t0})+K-\lambda  x^2 \leq 0, \forall x
\end{flalign}
and,
\begin{flalign}
\label{kkt2}
(1-p) h(x;F_{t0})+K-\lambda  x^2 = 0, \forall x \in \mathcal{S}_0,
\end{flalign} 
where $K=ph(0;F_{t0})-h(N)-I(F_{to}) + \lambda \beta_p$, $\lambda \geq 0$ is the Lagrangian multiplier and $\mathcal{S}_0$ is the support set of the optimum distribution. 

The capacity achieving distribution is discrete and can be proved using the techniques provided in \cite{peak1} and is omitted for brevity. The key steps of the proof include:
\begin{itemize}
\item{Identify the function $Q(x)$ which gives a necessary and sufficient condition for optimality.}
\item{Show that $Q(x)$ has an analytic extension $Q(z)$ over the whole complex plane.}
\item{Prove by contradiction that the zero set of  $Q(z)$ cannot have limit points in its domain of definition and is at most countable.}
\end{itemize}
Since any mass point $x$ of the optimum distribution function satisfies the condition $Q(x)=0$ the number of mass points of the optimum distribution is at most countable. 

Hence we find that the optimum input distribution  is not Gaussian. To get further insight, consider $\{B_k\}$ to be $iid$ binary random variables with $P[B_1=0]=p=1-P[B_1=1]$ and let $\{G_k\}$ be $iid$ random variables with distribution $F$. Then $X_k'=B_kG_k$ is the capacity achieving $iid$ sequence. Also,
\begin{eqnarray}
I(X_k';X_k'+N_k)&=& h(B_kG_k+N_k)-h(N_k)\nonumber\\
&=& h(B_kG_k+N_k)-h(B_kG_k+N_k|B_k)+h(B_kG_k+N_k|B_k)-h(N_k)\\\nonumber
&=& I(B_k;B_kG_k+N_k)+I(G_k;B_kG_k+N_k|B_k)\nonumber\\
&=& I(B_k;B_kG_k+N_k)+ (1-p)I(G_k;G_k+N_k).\label{newc}
\end{eqnarray}
This representation suggests the following interpretation (and coding theoretic implementation) of the scheme: the overall code is a superposition of a binary ON-OFF code and an $iid$  code with distribution $F$. The position of the ON (and OFF) symbols is used to reliably encode $I(B;BG+N)$ bits of information per channel use, while the code with distribution $F$ (which is used only during the ON symbols) reliably encodes $(1-p)I(G;G+N)$ bits of information per channel use.

It is interesting to compare this result with the capacity in \cite{meda}. The capacity result in \cite{meda} is only the second term in \eqref{newc} evaluated with $G_k$ being Gaussian.

\begin{figure}[ht]
\centering
\includegraphics [height=3 in, width= 3.5in ]{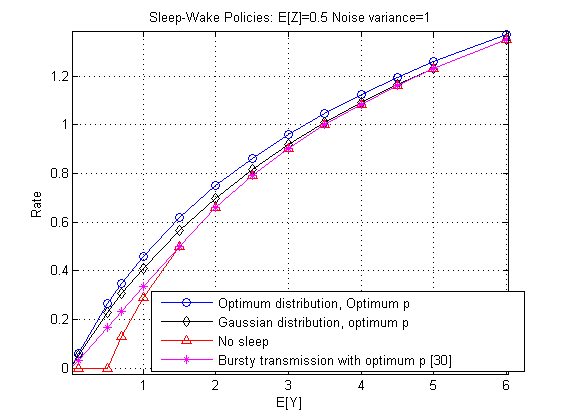}
\caption{Comparison of Sleep Wake policies}
\label{figmeda}
\end{figure}

In Fig.\ref{figmeda} we compare the optimal sleep-wake policy, a sleep wake policy with $F$ being mean zero Gaussian with variance $E[Y]/(1-p)-\alpha$  and no-sleep policy with the result in \cite{meda}. We take $E[Z]=0.5$ and $\sigma^2=1$.
We see that when $E[Y]$ is comparable or less than $E[Z]$ then the node chooses to sleep with a high probability. When $E[Y]>>E[Z]$ then the node will not sleep at all. Also it is found that when $E[Y]<E[Z]$, the capacity is zero when the node does not have a sleep mode. However we obtain a positive capacity if it is allowed  to sleep. When $E[Y]>>E[Z]$, the optimal distribution $F$ tends to a Gaussian distribution with mean zero and variance $E[Y]-\alpha$.

From the figure we see that our scheme improves the capacity provided in \cite{meda}. This is due to the embedded binary code and the difference is significant at low values of $E[Y]$. 

\section{Achievable Rate with Energy Inefficiencies}
\label{opt}
In this section we make our model more realistic by taking into account the
inefficiency in storing energy in the energy buffer  and the leakage from the energy buffer (\cite{jiang})
 for HSU architecture. For simplicity, we will ignore the energy $Z_k$ used for sensing and processing.

We assume that if energy $Y_k$ is harvested at time $k$, then only energy $\beta_1Y_k$ is stored in the buffer
  and  energy $\beta_2$ gets leaked in each slot where $0<\beta_1 \le 1$ and $0<\beta_2<\infty$.
 Then \eqref{eqn2} become
\begin{eqnarray}
E_{k+1}  = ((E_k - T_k)-\beta_2)^{+} + \beta_1Y_k. \label{eqne1}
\end{eqnarray}
The energy can be stored in a supercapacitor and/or in a battery. For a supercapacitor,
$\beta_1 \ge 0.95$ and for the Ni-MH battery (the most commonly used battery) $\beta_1 \sim 0.7$.
 The leakage  $\beta_2$ for the  battery is close to  0 but for the super capacitor it may be somewhat larger.

In this case, similar to the achievability of Theorem 1 we can show that
 \begin{eqnarray}
 R_{HSU} = 0.5~\log\left(1+\frac{\beta_1 E[Y]-\beta_2}{\sigma^2}\right) \label{eqnr1}
 \end{eqnarray}
 is achievable.
This policy is neither capacity achieving nor throughput optimal \cite{vinod1}.
An achievable rate of course is \eqref{fin} (obtained via HU). Now one does not even store energy and $\beta_1,~ \beta_2$ are not effective. The upper bound  $0.5~\log(1+{E[Y]} / {\sigma^2})$ is achievable if  $Y$ is chi-square distributed with degree 1. Now, unlike in Section III, the rate achieved by the HU may be larger than \eqref{eqnr1} for certain range of parameter values and distributions.

 Another achievable policy for the system with  an energy buffer   with storage inefficiencies is to use the harvested energy $Y_k$ immediately instead of storing in the buffer. The remaining energy after transmission is stored in the buffer. We call this \emph{Harvest-Use-Store} (HUS) architecture. For this case, \eqref{eqne1} becomes
  \begin{eqnarray}
E_{k+1}  = ((E_k + \beta_1(Y_k-T_k)^{+}- (T_k-Y_k)^{+})^{+}-\beta_2)^+. \label{eqne2}
\end{eqnarray}
Compute the largest constant $c$ such that $ \beta_1 E[(Y_k-c)^{+}] >  E[(c-Y_k)^{+}]+
 \beta_2$. This is the largest $c$ such that taking $E[T_k] < c$
will make $E_k \to \infty~a.s.$ %Also no (asymptotically) stationary
%ergodic $\{T_k\}$ with $E[T_k]>c$ is sustainable. 
Thus, as in
Theorem 1, we can show that rate
 \begin{eqnarray}
 R_{HUS} = 0.5~\log\left(1+\frac{c}{\sigma^2}\right) \label{eqnc1}
 \end{eqnarray}
 is achievable for this system. This is achievable by an input with distribution $iid$ Gaussian with mean zero and variance $c$.

Equation \eqref{eqne1} approximates the system where we have only rechargable battery while \eqref{eqne2} approximates the system where the harvested energy is first stored in a supercapacitor and after initial use transferred to the battery.

When  $\beta_1=1, \beta_2=0$ we have obtained the capacity of this system in Section III. For the general case, its capacity is an open problem.% although we conjecture that \eqref{eqnc1} may be the capacity.

We illustrate the achievable rates mentioned above via an example.

\begin{figure}[h]
\centering
\includegraphics [height=3 in, width= 3.5 in ]{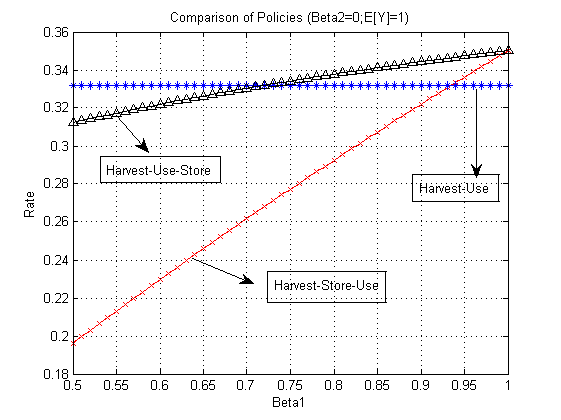}
\caption{Rates for various architectures}
\label{figineff}
\end{figure}

{\bf{Example 1}}\\
Let $\{Y_k\}$   be $iid$ taking values in  $\{0.25,0.5,0.75,1\}$ with equal probability. We take the
 loss due to leakage, $\beta_2=0$. In Figure \ref{figineff} we compare the various architectures discussed in this section for
varying storage efficiency $\beta_1$. We use the result in \cite{peak3} for computing the capacity in \eqref{fin}.
From the figure it can be seen that if the storage efficiency is very poor it is better to use the $ HU$ policy. This requires
no storage buffer and has a simpler architecture. If the
storage efficiency is good $HUS$ policy gives the best performance.
For $\beta_1=1$, the $HUS$ policy and
$HSU$ policy have the same performance. 
Thus if we judiciously use a combination of a supercapacitor and a battery, we may obtain a better performance.
%%%%%%%%%%%%%%%%%%%% START OF FADING
\section{Fading AWGN Channel}
\label{model1} 

\begin{figure}[h]
\begin{center}
\includegraphics[height=1.6in, width=4in]{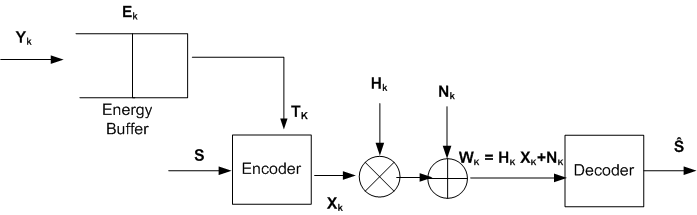}
\caption{The model} \label{fig1_1}
\end{center}
\end{figure}

In this section we extend the results of Theorem 1 to include fading. Rest of the  notation is same as in Section \ref{stability}. The model considered is given in Figure \ref{fig1_1}.

The encoder receives a message $S$ from the node and generates an $n$-length codeword to be transmitted on the fading AWGN channel. We assume flat, fast,  fading. At time $k$ the channel gain is $H_k$ and takes values in $\mathcal{H}$. The sequence $\{ H_k \}$ is assumed $iid$, independent of the  energy generation sequence $\{ Y_k \}$. The channel output at time $k$  is $W_k=H_k X_k+N_k$ where $X_k$ is the channel input at time $k$ and  $\{N_k\}$ is  \emph{iid} Gaussian noise with zero mean and variance $\sigma^2$. The decoder receives $Y^n \stackrel{\Delta}{=}(Y_1,...,Y_n)$ and reconstructs  $S$ such that the probability of decoding error is minimized. Also, the decoder has perfect knowledge of the channel state $H_k$ at time $k$.

If the channel input $\{X_k\}$ is AMS ergodic, then it can be easily shown that for the fading AWGN channel $\{(X_k,W_k),~k \ge 0\}$ is also AMS ergodic. Thus the channel capacity of the fading system is (\cite{gray})
\begin{equation}
\label {eqnn112}
C= \sup_{p_x} \overline{I}(X;W)= \sup_{p_x} ~ \limsup_{n \to \infty}\frac{1}{n} I(X^n,W^n)
\end{equation}
where under $p_x$, $\{X_n\}$ is an AMS sequence, $X^n=(X_1, ..., X_n)$ and the supremum is over all possible AMS sequences $\{X_n\}$. For a fading  AWGN channel, capacity achieving $X_k$ is zero mean Gaussian with variance $T_k$ where $T_k$ depends on  the power control policy used and is assumed AMS. Then  $E[T] \le E[Y]$ where $E[T]$ is the mean of $T$ under its stationary mean. The following theorem shows that one can find a sequence of codeword�s with code length $n$ and rate $R$ such that  the average probability of error goes to zero as $n \to \infty$ if and only if $R<C$ where $C$ is given in \eqref{faeq}.

{\bf Theorem 3} For the energy harvesting system with perfect CSIT, 
\begin{equation}
 C= 0.5~E_{H}\left[\log(1+\frac{H^2T^*(H)}{\sigma^2})\right],
 \label{faeq}
 \end{equation}
where
\begin{equation}
T^* (H) = \left(\frac{1}{H_0}-\frac{1}{H}\right)^+,\label{fto} 
\end{equation}
and $H_0$ is chosen such that $E_H[T^*(H)] = E[Y]$.

{\bf Proof:}  See Appendix C.
\vspace{0.2cm}

Thus we see that the capacity of this fading channel is same as that of a node with average power constraint $E[Y]$ and the instantaneous power allocated is according  to  `water filling' power allocation. The hard energy constraint of $E_k$ at time $k$ does not affect its capacity.
The capacity achieving signaling for our system is  $X_k=sgn(X_k')\min(\sqrt{T^*(H_k)} |X_k'|,\sqrt{E_k})$, where $\{X_k'\}$ is $iid$ $\mathcal{N}(0,1)$ and $T^*(H)$ is defined in \eqref{fto}. 

When  no CSI is available at the transmitter (but perfect CSI is available at  the decoder), take  $X_k=sgn(X_k')\min( |X_k'|,\sqrt{E_k})$ where $\{X_k'\}$ is $iid$ $\mathcal{N}(0,E[Y])$ and as in Theorem 1 this approaches the capacity of  $0.5~E_H[\log(1+{H^2 E[Y]}/{\sigma^2})]$. 

Similar to the non-fading case the throughput optimal policies in \cite{vinod1} are  related to the 
 Shannon capacity achieving energy management policies provided here for  the infinite buffer case.
Also the capacity is the same as the maximum throughput obtained  in the data-buffer case in \cite{vinod1}.

If there is no energy buffer to store the harvested energy then at time $k$ only $Y_k$ energy is available. Thus $X_k$ is peak power limited to $Y_k$. The capacity achieving distribution for an AWGN channel with peak power constraint $Y_k=y$  is not Gaussian. Let $X(y,\sigma^2)$ be a random variable with the capacity achieving distribution for an AWGN channel with peak power constraint $y$ and noise variance $\sigma^2$. In general this distribution is discrete. Thus, if CSIT is exact then the transmitter will transmit $X(y,\sigma^2/h^2)$ at time $k$ when $Y_k=y$ and $H_k=h$. Therefore the ergodic capacity with  $Y_k$ information being available at the receiver is $0.5 E_{YH}[I(X(Y,\sigma^2/H^2);W)]$. If there is no CSIT then we can transmit $X(y,\sigma^2)$ and the  corresponding capacity is   $0.5 E_{YH}[I(X(Y,\sigma^2);W)]$.

 \subsection{ Capacity with Energy Consumption in Sensing and Processing}
\label{alloc}
In this section we extend the results in Section \ref{simulation} to the fading case.

First we extend the achievable policies given above to incorporate the energy consumption in activities other than transmission. We assume perfect CSIR for the channel state $H_k$ at the time $k$.  When there is perfect CSIT also, we use the signaling scheme   $X_k=sgn(X_k')\min(\sqrt{T^*(H_k)} |X_k'|,\sqrt{E_k})$, where $\{X_k'\}$ is $iid$ $\mathcal{N}(0,1)$ and $T^*(H)$ is the optimum power allocation such that  $E[T^*(H)]= E[Y]-E[Z]-\epsilon$. When  no CSI is available at the transmitter, we use $X_k=sgn(X_k')\min( |X_k'|,\sqrt{E_k})$ where $\{X_k'\}$ is $iid$ $\mathcal{N}(0,E[Y]-E[Z]-\epsilon)$. The achievable rates  for CSIT and no CSIT  respectively are,

\begin{gather}
R_{PE-CSIT}= 0.5~E_H\left[\log\left(1+\frac{H^2 T^*(H)}{\sigma^2}\right)\right],\label{abcd}\\
R_{PE-NCSIT}= 0.5~E_H\left[\log\left(1+\frac{H^2(E[Y]-E[Z]-\epsilon)}{\sigma^2}\right)\right].\label{abc1}
\end{gather}

When Sleep Wake modes are supported  the achievable rates can be improved as in Section \ref{simulation}. 

{\bf Theorem 4} Let $\mathcal{P}(H)$ be the set of all feasible power allocation policies such that for $P(H) \in \mathcal{P}(H)$, $E_H[P(H)] \le E[Y]$.  For the energy harvesting system with processing energy transmitting over a fading Gaussian channel,

\begin{equation}
C= \sup_{P(H) \in \mathcal{P}(H)} ~ \sup_{p_x: E[b(X)] \le P(H)} E[I(X;W)] \label{impee1}
\end{equation}
is the capacity for the system. 

{\bf Proof:}  : See Appendix D.
\vspace{0.2cm}

We  compute the capacity \eqref{impee1} and the capacity achieving distribution. Let $P^*(h)$ be the power allocated in state $h$. Without loss of generality, under $H=h$, the node sleeps with probability $p, (0 \leq p \leq 1)$ and with probability $(1-p)$ the node transmits with a distribution $F_t(.)$.  As in Section  IV, we can show using KKT conditions that the capacity achieving distribution for state $H=h$ is discrete and the number of mass points are at most countable with  $E[b(X)] \le P(h)$. As in the case without fading the distribution $F_t(.)$ under $H=h$ is not Gaussian.

The optimal power allocation policy $P^*(H)$ that maximizes \eqref{impee1} is  not 'water filling' but similar and uses more power when the channel is better.

{\bf{Example 2}}

\begin{figure}[ht]
\centering
\includegraphics [height=3in, width=3.5in ]{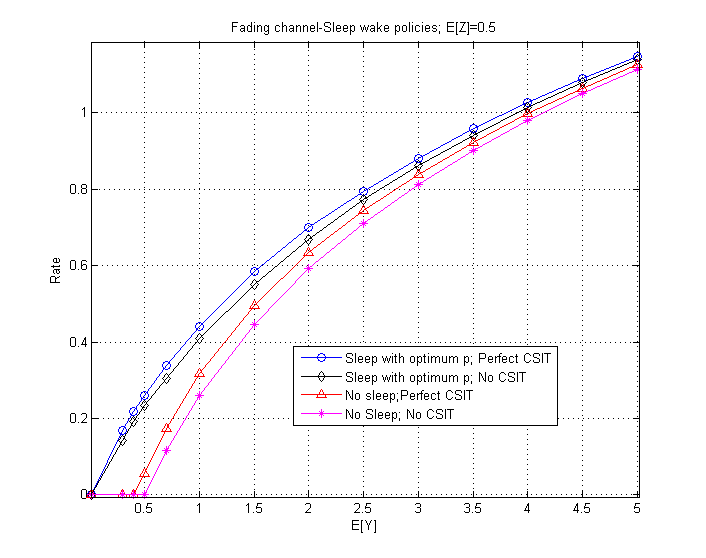}
\caption{Comparison of sleep wake policies}
\label{figsleep1}
\end{figure}

Let the fade states take values in $\{0.5,~1,~1.2\}$ with probabilities $\{0.1,~0.7,~0.1\}$. We  take $\alpha=E[Z]=0.5,~ \sigma^2=1$. We compare the capacity for the cases with perfect and no CSIT when there is no sleep mode supported (Equation \eqref{abcd}, \eqref{abc}) and with the optimal sleep probability in Figure \ref{figsleep1}.

From the figure we observe that
 \begin{itemize}
	\item The randomized sleep wake policy improves the rate significantly when $E[Y] \leq E[Z]$.
	\item The sensor node chooses not to sleep when $E[Y] >> E[Z]$.
\end{itemize}

\subsection{Achievable Rate with Energy Inefficiencies}
\label{opt1}
In this section we take into account the inefficiency in storing energy in the energy buffer  and the leakage from the energy buffer. The notation is same as in Section \ref{opt}. 

The energy evolves as
\begin{eqnarray}
E_{k+1}  = ((E_k - T_k)-\beta_2)^{+} + \beta_1Y_k. \label{eqne11}
\end{eqnarray} 
In this case, similar to the achievability of Theorem 3 we can show that the  rates 
\begin{eqnarray}
 R_{S-NCSIT} = 0.5~E_H\left[\log\left(1+\frac{H^2(\beta_1 E[Y]-\beta_2)}{\sigma^2}\right)\right], \label{eqnr11}\\
R_{S-CSIT} = 0.5~E_H\left[\log\left(1+\frac{H^2(\beta_1 T(H)-\beta_2)}{\sigma^2}\right)\right],\label{eqnr2}
 \end{eqnarray}
are achievable in the no CSIT and perfect CSIT  case respectively, where $T(H)$ is a power allocation policy such that \eqref{eqnr2} is maximized subject to $E_H[T(H)]\le E[Y]$. This policy is neither capacity achieving nor throughput optimal \cite{vinod1}.

An achievable rate when there is no buffer and perfect CSIT is
 \begin{eqnarray}
C= E_{YH}[I(X(Y,H);W)],\label{eqnr3}
 \end{eqnarray}
where $X(y,h)$ is the distribution that maximizes the capacity subject to peak power constraint $y$ and fade state $h$. A numerical method to evaluate the capacity with peak power constraints is provided in \cite{peak1}. It is also shown in \cite{peak3} that for $\sqrt{y} <1.05$, the capacity has a closed form expression 

\begin{eqnarray}
C(y)= y-\int_{-\infty}^{\infty} \frac{e^{-x^2/2} \log {cosh(y-\sqrt{y}x)}}{\sqrt{2\pi}} dx. \label{eqnrr3}
\end{eqnarray}

When there is no buffer and no CSIT the distribution that maximizes the capacity cannot be chosen as in \eqref{eqnr3} and the capacity is less than the capacity given in \eqref{eqnr3}. The capacity in  \eqref{eqnr3}  is without using buffer and hence $\beta_1$ and $\beta_2$ do not affect the capacity. Hence unlike in  Section III, \eqref{eqnr3} may be larger than \eqref{eqnr11} and \eqref{eqnr2} for certain range of parameter values. We will illustrate this by an example.
 
For the  \emph{Harvest-Use-Store} (HUS) architecture,  \eqref{eqne11} becomes
\begin{eqnarray}
E_{k+1}  = ((E_k + \beta_1(Y_k-T_k)^{+}- (T_k-Y_k)^{+})^{+}-\beta_2)^{+}. \label{eqne21}
\end{eqnarray} 
Find the largest constant $c$ such that $ \beta_1 E[(Y_k-c)^{+}] \ge  E[(c-Y_k)^{+}]+ \beta_2$. Of course $c<E[Y]$. When there is no CSIT, this is the largest $c$ such that taking $T_k= \min(c-\delta,E_k)$, where $\delta >0$ is any small constant,  will make $E_k \to \infty~a.s.$ and hence $T_k \to c~a.s.$ Then, as in Theorem 3, we can show that
 \begin{eqnarray}
 R_{US-NCSIT} = 0.5~E_H\left[\log\left(1+\frac{H^2c}{\sigma^2}\right)\right] \label{eqnc11}
 \end{eqnarray}
 is an achievable rate.
 
 When there is perfect CSIT, 'water filling' power allocation can be done subject to average power constraint of $c$ and the achievable rate is
  \begin{eqnarray}
 R_{US-CSIT} = 0.5~E_H\left[\log\left(1+\frac{H^2T^{*}(H)}{\sigma^2}\right)\right] \label{eqnc2}
 \end{eqnarray}
  where $T^{*}(H)$ is the 'water filling' power allocation with $E[T^{*}(H)] = c$.

We illustrate the achievable rates mentioned above via an example.

{\bf{Example 3}}

\begin{figure}[!h]
\centering
\includegraphics [height=3 in, width=3.5in ]{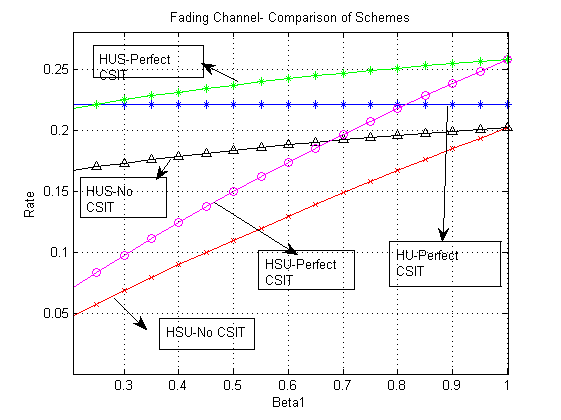}
\caption{Rates for various architectures}
\label{figineff1}
\end{figure}

Let the process $\{Y_k\}$   be $iid$ taking values in  $\{0.5,~ 1\}$ with probability $\{0.6,~  0.4\}$ . We take the
 loss due to leakage $\beta_2=0$. The fade states are $iid$ taking values in $\{0.4, ~0.8,~1\}$ with probability $\{0.4,~ 0.5,~ 0.1\}$. In Figure \ref{figineff1} we compare the various architectures discussed in this section for varying storage efficiency $\beta_1$. The capacity for the no buffer case with perfect CSIT is computed using equations \eqref{eqnrr3} and \eqref{eqnr3}.

From the figure we observe

\begin{itemize}
	\item Unlike the ideal system, the $HSU$ (which uses infinite energy buffer) performs worse than the $HU$  (which uses no energy buffer) when storage efficiency is poor for the perfect CSIT case.
	\item When storage efficiency is high, $HU$ policy performs worse compared to $HSU$ and $HUS$ for perfect CSIT case.
	\item $HUS$ performs better than $HSU$ for No/Perfect CSIT.
	\item For $\beta=1$, the $HUS$ policy and $HSU$ policy are the same for both perfect CSIT and no CSIT.
	\item The availability of CSIT and storage architecture plays an important role in determining the achievable rates.
\end{itemize}

 \section{Combining Information and Queuing Theory}
\label{CIQT}
\begin{figure}
\begin{center}
\includegraphics[height=1.4in, width= 3.4 in ]{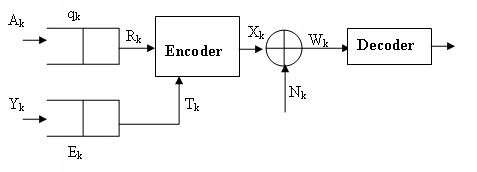}
\caption{Model of an energy harvesting point to point channel.}
\label{fffig1}
\end{center}
\end{figure}

In this section we consider a system with both energy and data buffer, each with infinite capacity (see Fig. \ref{fffig1}). We consider the simplest case: no fading, no battery leakage and storage inefficiencies.  The system is slotted. During slot $k$ (defined as time interval $[k, k+1]$, i.e., a slot is a unit of time), $A_k$ bits are generated. Although the transmitter may generate data as packets, we allow arbitrary fragmentation of packets during transmission. Thus, packet boundaries are not important and we consider bit strings (or just fluid). The bits $A_k$ are eligible for transmission in $(k+1)$st slot. The queue length (in bits) at time $k$ is $q_k$. We assume that transmission consumes most of the energy at the transmitter and ignore other causes of energy consumption.  We denote by $E_k$ the energy available at the node at time $k$. The energy harvesting source  is able to replenish energy by $Y_k$ in slot $k$.

In slot $k$ we will use energy \begin{equation} T_k = \min(E_k, E[Y]-\epsilon), \label{disp11} \end {equation} where $\epsilon$ is a small positive constant. It was shown in \cite{vinod1} that such a policy is throughput optimal (and it is capacity achieving in Theorem 1).

There are $n$ channel uses (mini slots) in a slot, i.e., the system uses an $n$ length code to transmit the data in a slot. The length $n$ of the code word can be chosen to satisfy certain code error rate. The slot length $n$ and $R_k$ are to be appropriately chosen.   We use codewords of length $n$ and rate $R_k=0.5 \log(1+ T_k/{n \sigma^2})$ in  slot $k$  with the following coding and decoding scheme:

1) An augmented message set $\{1,...,2^{nR_k}\} \cup \{0\}$.

2) An encoder that assigns a codeword $x^n (m)$ to each $m \in \{1,...,2^{nR_k}\} \cup \{0\}$ where $x^n(m)$ is generated as an $iid$ sequence with distribution $\mathcal{N}(0,T_k/n-\delta_1)$ and  $\delta_1 >0$ is a small constant. The codeword $x^n(m)$ is retained if it satisfies the power constraint $\sum_i^n x_i^2 \le T_k$. Otherwise error message 0 is sent.

3) A decoder that assigns a message $\hat{m} \in \{1,...,2^{nR_k}\} \cup \{0\}$  to each received sequence $w^n$ in a slot such that $(x^n(\hat{m}),w^n)$ is jointly typical and there is no other $x^n(m')$ jointly typical with $w^n$. Otherwise it declares an error.

In slot $k$, $n R_k$ bits are taken out of the queue if $q_k \geq nR_k$. The bits are represented by a message $m_k \in \{1,...,2^{nR_k}\}$ and $x^n(m_k)$ is sent. If $q_k  < nR_k$ no bits are taken out of the queue and ``0 message" $x^n(0)$ is sent.

Hence the processes $\{E_k\}$ and $\{q_k\}$  satisfy
\begin{eqnarray}
q_{k+1} &=&  q_k -nR_kI_{\{q_k \ge nR_k\}} + A_k,\\ \label{eqn1111}
E_{k+1} & = & (E_k - T_k) + Y_k. \label{eqn2222}
\end{eqnarray} 

 With  $T_k$ in \eqref{disp11}, $E_k \to \infty~a.s.$ and  $T_k \to E[Y]-\epsilon~a.s.$  Also, $R_k = 0.5~\log(1+\frac{T_k}{n \sigma^2}) \to 0.5~\log(1+\frac{E[Y]-\epsilon}{n \sigma^2})$.
Thus we obtain

{\bf Theorem  5.} The random data arrival process $\{A_k\}$ can be communicated with arbitrarily low average probability of block error,  by an energy harvesting sensor node over a Gaussian channel   with a stable queue if and only if   $ E[A] < 0.5 n ~\log(1+\frac{E[Y]}{n \sigma^2}) $.~~~~~~~~~~~~~~~~~~~~~~~~~~~~~~~~~~~~~~~~~~~~~~~~~~~~~~~~~~~~~~~~~~~~~~~~~~$\blacksquare$

In Theorem 5  'stability' of the queue has the following interpretation. If $\{A_k\}$ is stationary, ergodic then $P[q_k \to \infty]=0$ and with probability 1,  $\{q_k\}$  visits the set $\{q: q< nR\}$ infinitely often.  Also the sequence $\{q_k\}$ is tight (\cite{bil}). If $\{A_k\}$ is iid then $\{q_k,E_k\}$ is a Markov chain. With $T_k$ in \eqref{disp11}, asymptotically, $T_k \to E[Y]-\epsilon~a.s.$ and we can ignore the $E_k$ component of the process and think of $\{q_k\}$ as a Markov chain with $T_k= E[Y]-\epsilon$. It has a finite number of ergodic sets. The process  $\{q_k\}$ eventually enters one ergodic set with probability 1 and then approaches a stationary distribution. If $\{q_k\}$ is irreducible and aperiodic then $\{q_k\}$ has a unique stationary distribution and $\{q_k\}$ converges in distribution to it irrespective of initial conditions.

Although the capacity achieved in each slot is as per Theorem 1, the set-up used here is somewhat different. In  Theorem 1, the time scale of the dynamics of the energy process $\{E_k\}$ is mini slots, but in this section we have taken it at the time scale of slots (which one is the right model depends on the system under consideration). Thus, in  Theorem 1 we used the theoretical tool of AMS sequences. But in our present setup, in  a slot we can use $X_1, X_2,....,X_n$  $iid$ Gaussian $\mathcal{N}(0,T_k/n-\delta)$ and use a codeword only if it satisfies $X_1^2+....+X_n^2 \le T_k$ and $q_k \geq nR_k$; otherwise  an error message is sent. Of course, if the physical system demands that we should use for the energy dynamics the time scale of a channel use then we can  use the framework of Theorem 1.
%%%%%%%%%%%%%%%%%END OF FADING
\section{Finite Buffer}

In this section we find achievable rates when the sensor node has a finite buffer to store the harvested energy. This case is of  more practical interest.  We consider the simplest case: no fading, no battery leakage and storage inefficiencies and no data queue.  The node  has an energy buffer  of size $\Gamma < \infty$.  By this we mean that the energy buffer can store a finite number of energy units of interest.

We use the HUS architecture where the energy harvested is used  and only the left over energy is stored. The energy available at the buffer at time $k$ is denoted by $\hat{E}_k$. At time $k$, the node uses energy $T_k$ with $T_k \le \hat{E}_k+Y_k \stackrel{\Delta} {=} E_k$.  We assume that  $\hat{E}_k$ and $Y_k$  take values in finite alphabets.   Also,  $\{Y_k\}_{k \geq 1}$ is assumed $iid$. 

%(see for example \cite{rao})

We assume that the buffer state information (BSI), ${E_k}$, is perfectly available at  the encoder and the decoder at time $k$. $X_k$ denotes the codeword symbol used at time $k$ and $X_k^2\le T_k$. Of course $T_k \le E_k$ and $\hat{E}_k \le \Gamma$.  In general $T_k$ is a function of $E_0,...,E_k$. . An easily tractable class of energy management policies is 
\begin{eqnarray}
\label{abccc}
T_k = h(E_k),
\end{eqnarray}
where $h$ defines the energy management policy. The codeword symbol  $X_k$ is picked with a distribution that maximizes the capacity of a Gaussian channel with peak power constraint $T_k$ (we quantize this such that $\{E_k\}$ takes values in a finite alphabet). Hence the process $\{E_k\}_{k \geq 1}$ satisfies,  
%{\bf{this maximizes only one point, we need to give for the whole rate region}}
 
\begin{equation}
E_{k+1}=(E_k-X_k^2) + Y_{k+1}
\end{equation}
and is a finite state Markov chain with the transition matrix decided by $h$.  If $\hat{E}_0=0$ then the Markov chain will either enter only one ergodic set or possibly in a finite number of disjoint components which depend on $h$. If $I^*$ and $\overline{I}$ denote the Pinsker and Dobrushin information rates (\cite{gray}),  since we have finite alphabets, $I^{*}= \overline{I}$. In particular,
\begin{align} 
\label{aabb}
I^{*}\big(X;W \big) {= \overline{I}\big(X;W\big)} {= \lim_{n \rightarrow \infty}~\frac{1}{n}I(X^{(n)};W^{(n)}).}
\end{align}
  Also, Asymptotic Equi-partition Property holds for $\{X_k,W_k\}$.

%For definitions of $I^{*}= \overline{I}$, (See \cite{gray}).
The following theorem  provides achievable rates.

{\bf{Theorem 6:}} A rate $R$ is achievable if an  energy management policy $h$ exists, such that  $R < \overline{I}\big(X;W)$.$\blacksquare$         

The proof is similar to the achievability proof given in Theorem 1. The rates \eqref{aabb} can be computed via algorithms available in \cite{sk} and \cite{vo}. Using stochastic approximation  (\cite{sa}) we can obtain the Markov chains that optimize \eqref{aabb}.
If initial energy  $\hat{E}_0$ is not zero, then the Markov chain can enter some other ergodic sets and the achievable rates can be different. If $h$ is such that $\{E_k\}$ is an irreducible Markov chain then the achievable rates will be independent of the initial state $\hat{E}_0$.

Theorem 6 can be generalized to include the case where $\{(E_k,X_k)\}$ is a $k-$step finite state Markov chain. In fact if $\{(E_k,X_k)\}$ is a general AMS ergodic finite alphabet sequence then AEP holds and $I^{*}=\overline{I}$. Thus, $R< \lim_{n \to \infty} n^{-1} I (X^n;W^n)$ is achievable.

 The capacity of our system can be written  as (\cite{hanverdu})
\begin{equation}
C= \sup~ p-\liminf \frac{1}{n} log \frac{p(x^n,w^n)}{p(x^n)p(w^n)},
\label{han}
\end{equation}
where $p-\liminf$ is defined in \cite{hanverdu} and $\sup$ is over all input distributions $X^n$ which satisfy the energy constraints $X_k^2 \le E_k$ for all $k \ge 0$. An interesting open problem is: can \eqref{han}  be obtained by limiting $\{X_n\}$  to AMS ergodic sequences mentioned above?

 The achievable rates when the  decoder has  only partial information about  $ E_{k}$ can be handled as for the system with no buffer and partial BSI, studied in  Section III.

{\bf{Example 4}}

\begin{figure}
\begin{center}
\includegraphics[height=2.5in, width= 3.5 in ]{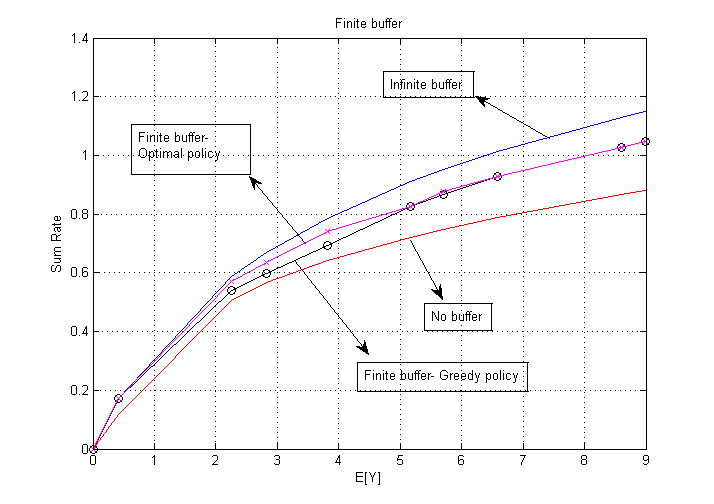}
\caption{Acheivable rate for finite buffer}
\label{f11}
\end{center}
\end{figure}

We consider a system with a  finite  buffer with $\Gamma = 15$ units in steps of  size 1. The $Y_k$ process has three mass points and  provided in Table 1. We compute the optimal achievable rate using simultaneous perturbation stochastic approximation algorithm \cite{sa}. The  achievable rate is also compared with a greedy policy,  where the rate is evaluated using algorithms provided in \cite{sk} and  \cite{vo}. In the greedy policy,  at any instant $k$, an optimum distribution for an AWGN channel peak amplitude constrained to $\sqrt{E_k}=\sqrt{\hat{E}_k+Y_k}$ is used. We have also obtained the optimal rates using a 1-step Markov policy \eqref{abccc} where the optimal Markov chain is obtained via stochastic approximation. Then achievable rates  are  compared with the capacity with infinite buffer and no-buffer in Figure \ref{f11}.

\begin{table}[h!]
\label{tab1}
\caption{$Y_k$ Process}
  \begin{center}
\begin{tabular} { |c | c || c |}
\hline
E(Y) & Mass points & Probabilities\\
\hline
0 &0~ 0~ 0 & 1~~~~~~~ 0~~~~~~~~ 0\\
\hline
1.0141 & 0 ~ 1~ 2 & 0.3192 ~ 0.3474~ 0.3333\\
\hline
2.1031 & 1 ~2~ 3 &0.2303~ 0.4364~ 0.3333\\
\hline
3.3078 &2 ~3 ~4 &0.1794~ 0.3333 ~0.4872\\
\hline
4.1990 &3 ~4~ 5 &0.2338~ 0.3333 ~0.4329\\
\hline
5.0854 &4 ~5 ~6 &0.3333~ 0.2479 ~0.4188\\
\hline
5.8738 &5~ 6 ~7 &0.3964~ 0.3333 ~0.2703\\
\hline
6.7168 &6 ~7 ~8 &0.4749~ 0.3333 ~0.1917\\
\hline
8.2533 &7~ 8~ 9 &0.2067~ 0.3333 ~0.4600\\
\hline
9.0332 &8~ 9 ~10 &0.3167~ 0.3333 ~0.3499\\
\hline
9.9136 &9 ~10 ~11 &0.3333 ~0.4198 ~0.2469\\
\hline

\end{tabular}
\end{center}
\end{table}

From the figure we observe that, for a given buffer size, the greedy policy is close to optimal at higher $E[Y]$. Also, the optimal achievable rates for finite buffer case are close to the capacity for infinite buffer for small $E[Y]$ but becomes close to the greedy   at high $E[Y]$.

\section{Conclusions}
\label{conclude}
In this paper  the Shannon capacity of  an energy harvesting sensor node transmitting over an AWGN Channel is provided. It is shown that the  capacity achieving policies are related to the throughput optimal policies. Also, the capacity is provided when energy is consumed in activities other than transmission. Achievable rates are  provided when there are inefficiencies in energy storage. We  extend the results to the fast fading case. We also combine the information theoretic and queuing theoretic formulations. Finally we also consider the case when the energy buffer is finite.

\section{Acknowledgement}
The authors would like to thank Deekshith P K, PhD student, ECE, IISc for helpful inputs provided in Section IV.

\appendices
\section{Proof of Theorem 1}
{\emph{Codebook Generation :}} Let $\{X_k'\}$ be an $iid$ Gaussian sequence with mean zero and
variance $E[Y]-\epsilon$  where $\epsilon > 0$ is an arbitrarily
small constant. For each message $s \in \{1,2,...,2^{nR}\}$,
generate $n$ length codewords according to the iid distribution
$\mathcal{N}(0,E[Y]-\epsilon)$. Denote the codeword by 
$X'^n(s)$. Disclose this codebook to the receiver.

{\emph{Encoding}}: When $S=s$, choose the channel codeword to be  $X_k= sgn (X_k'(s))
\min( \sqrt{E_k},|X_k'(s)|)$ where $sgn(x)=1$ if $x\ge0$ and $=-1$
if $x<0$. Then $T_k=X_k^2 \le E_k$ and $E[T_k]=E[X_k^2]\le
E[Y]-\epsilon$. Thus, from standard results on G/G/1 queues
(\cite{gg1},~chapter 7) $E_k \to \infty~a.s.$ and hence $|X_k-X_k'|
\to 0~a.s$. Also $(X_k,W_k)$ converges almost surely(a.s.) to a random variable with the distribution of  $(X_k',W_k')$ and  $\{(X_k,W_k,X_k')\}$ is AMS ergodic  where $W_k'=X_k'+N_k$.

{\emph{Decoding:}} The decoder obtains $W^n$ and finds the  codeword $X'^n(\hat{s})$ such that $(X'^n(\hat{s}),W^n) \in T_{\epsilon}^n$ where $T_{\epsilon}^n$ is the set of  weakly $\epsilon$-typical sequences of the joint AMS ergodic distribution $P_{X'W'}$. If it is a unique $\hat{s}$ then it declares $\hat{s}$ as the message transmitted; otherwise declares an error.

{\bf{Analysis of error events}}

Let $s$ has been transmitted. The following error events can happen

{\bf{E1}}: $\{(X'^n({s}),W^n) \notin T_{\epsilon}^n\}$. The probability of 
event {\bf{E1}} goes to zero as,   $\{X'_k, W_k\}$ 
 is AMS ergodic and AEP holds for AMS ergodic sequences  (\cite{bar}), as $\{X'_k, W_k\}$  has a density with respect to $iid$ Gaussian measure on an appropriate Euclidean space.

{\bf{E2}}:  There exist $\hat{s} \ne s$ such that
$\{(X'^n(\hat{s}),W^n) \in T_{\epsilon}^n\}$. Let $\overline{H}(X'),\overline{H}(W')$ be the entropy rates of $\{X'_k\}$ and $\{W'_k\}$. Next we show that $P(E_2) \to 0$ as $n \to \infty$. We have

\begin{eqnarray*}
\label{eqne3}
P(E2)&=& \sum_{\hat{s} \ne s} P((X'^n(\hat{s}),W^n) \in T_{\epsilon}^n)\\
&\le& 2^{-nR} \sum_{(x^n,w^n) \in T_\epsilon^n} P(x^n,w^n)\\
&\le& 2^{-nR} \sum_{(x^{'n},w'^n) \in T_\epsilon^n} P(x'^n) P(w'^n)\\
&\le& |T_{\epsilon}^n|2^{-nR}2^{-(n\overline{H}(X')-\epsilon)} 2^{-(n\overline{H}(W')-\epsilon)} \\
&\le& 2^{(n\overline{H}(X',W')+ \epsilon)} 2^{-nR}
2^{-(n\overline{H}(X')-\epsilon)} 2^{-(n\overline{H}(W')-\epsilon)}.
\end{eqnarray*}

Therefore, $P(E2) \to 0$ and $n \to \infty$ if $R <
\overline{I}({X'};{W'})=0.5\log(1+P/\sigma^2)$.

\emph {Converse  Part}:  For the system under consideration
 $\frac{1}{n} \sum_{k=1}^n T_k \le \frac{1}{n} \sum_{k=1}^n Y_k \to E[Y] ~a.s.$
Hence, if $\{X_k(s),~k=1,...,n\}$ is a codeword for message $s \in \{1,...,2^{nR}\}$ then for
 all large $n$ we must have $\frac{1}{n} \sum_{k=1}^n X_k(s)^2 \le E[Y]+\delta$ with a large probability for any $\delta >0$.
Hence by the converse in the AWGN channel case,
 $\limsup_{n \to \infty} \frac{1}{n} I(X^n;W^n)\le 0.5~\log(1+{(E[Y]+\delta)}/{\sigma^2})$. Now take $\delta \to 0$.

Combining the direct part and converse part completes the proof. ~~~~~~~~~~~~~~~~~~~~~~~~~~~~~~~~~~~~~~~~~~~~~~~~~~~~~~~~~~~~~~~~
 {\raggedleft{$\blacksquare$}}

\section{Proof of Theorem 2}\label{l1}

{\emph{Codebook Generation :}}  For each message $s \in \{1,2,...,2^{nR}\}$, generate $n$ length codewords according to an  $iid$ distribution $p_X'$ with constraint $E[b(X')]=E[Y]-\epsilon$, where $ \epsilon>0$ is a small constant. Denote the codeword by 
$X'^n(s)$. Disclose this codebook to the receiver.

{\emph{Encoding}} : When $S=s$, choose the channel codeword as 

\begin{eqnarray*}
X_k(s)= \begin{cases}
\min\{X_k'(s), \sqrt{(E_k-Z_k)^{+}}\},~& \text{if $X_k'\ge 0$},\\
\max\{X_k'(s), -\sqrt{(E_k-Z_k)^{+}}\},~& \text{if $X_k'< 0$}.
\end{cases}
\end{eqnarray*}
Then to transmit $X_k(s)$ we need energy $T_k= (X_k^2+Z_k)1_{\{X_k \ne 0\}}$ and $E_{k+1}= (E_k-T_k)+Y_k$. Also,
\begin{eqnarray*}
E[T_k]&=& E[X_k^2]+E[Z_k]P\{X_k \ne 0\}\\
&\le& E[X_k'^2]+\alpha P\{X_k' \ne 0\}= {E[Y]}-\epsilon.
\end{eqnarray*}

Thus, from standard results on G/G/1 queues
(\cite{gg1},~chapter 7) $E_k \to \infty~a.s.$ and hence $|X_k-X_k'|
\to 0~a.s$. Also finite dimensional distributions of $ \{(X_{m+k},W_{m+k},X_{m+k}'), ~k \ge 0\}$ converge $a.s.$ to that of $ \{(X_k',W_k',X_k')\}$. Thus $\{(X_k,W_k,X_k')\}$ is AMS ergodic with limiting distribution $(X_k',W_k',X_k')$ where $W_k'=X_k'+N_k$. Furthermore the energy constraints are also met.

If the chosen codeword is $\epsilon-$weakly typical and $\sum_{i=1}^n b(x_i)/n \le E[Y]-\epsilon$, then transmit it; otherwise send an error message. The probability that an error message is sent goes to zero as $n \to \infty$.

{\emph{Decoding:}} The decoder obtains $W^n$. If it  finds a unique codeword $X'^n(\hat{s})$ such that $\{(X'^n(\hat{s}),W^n) \in T_{\epsilon}^n\}$ where,  $T_{\epsilon}^n$ is the set of  $\epsilon$-typical sequence for the distribution $P_{X'W'}$, it declares $\hat{s}$ as the transmitted message. Otherwise it declares an error.

By the usual methods as in Theorem 1  with the above coding-decoding scheme and also  the fact that $C(.)$ is non-decreasing, we can show that the probability of error for this scheme goes to zero as $n \to \infty$. Thus we can achieve the capacity \eqref{impee}.

\emph{Converse}:
The converse follows via Fano's inequality as in Theorem 1. For that proof to hold here, we need that $C(.)$ is concave ~~~~~~~~~~~~~~~~~~~~~~~~~~~~~~~~~~~~~~~~~~~~~~~~~~~~~~~~~~~~~~~~~~~~~~~~~~~~~~~~~~~~~~~{\raggedleft{$\blacksquare$}}

\section{Proof of Theorem 3}
\emph{Achievability}: Let $T_k'=T^*(H_k)$ with $T^*$ defined in \eqref{fto} with $E[T^{*}(H)]=E[Y]-\epsilon$ where $\epsilon > 0$ is a small constant. Since $\{H_k\}$ is $iid$, $\{T_k'\}$ is also $iid$. We take $T_k= min(E_k,T_k')$. Thus, as in the proof of Theorem 1, from standard results on G/G/1 queues (\cite{gg1},~chapter 7) $E_k \to \infty~a.s.$ Therefore, as $T^*(H)$ is upper bounded, $\lim_{n \to \infty} \sup_{k \ge n} |T_k-T^*(H_k)| \to 0~a.s.$

 Let $\{X_k^{'}\}$ be $iid$ Gaussian with mean zero and variance one. The channel codeword $X_k= sgn(X_k')\min(\sqrt{T_k} |X_k'|,\sqrt{E_k})$ where $sgn(x)=1$, if $x \ge 0$ and $-1$ otherwise.  This is an AMS ergodic sequence with the stationary mean being the distribution of  $\sqrt{T^*(H_k)}X_k'$. Then since AWGN channel under consideration is AMS ergodic (\cite{gray}), $(X,W)\stackrel{\Delta}{=} \{(X_k,W_k), ~k \ge 1\}$ is AMS ergodic.   

By using the  techniques in Theorem 1, $ R \le \overline{I}(X,W)=0.5~E_H [\log(1+{H^2 T^{*}(H))}/\sigma^2)]$. 

\emph {Converse  Part}: Let there be a sequence of codebooks for our system with rate $R$ and average probability of error going to 0 as $n \to \infty$. If $\{X_k(s),~k=1,...,n\}$ is a codeword for message $s \in \{1,...,2^{nR}\}$ then  ${1}/{n} \sum_{k=1}^n X_k(s)^2 \le 1/n \sum_{k=1}^n Y_k \le E[Y] + \delta$ for any $\delta >0$  with a large probability for all $n$ large enough. 
Hence by the converse in the fading AWGN channel case (\cite{varia}),
 $R < \limsup_{k \to \infty}  I(X^k;W^k)/k \le  0.5~E_H[\log(1+{H^2 T^*(H)}/\sigma^2)]$ for $T^*(H)$ given in \eqref{fto}.

Combining the direct  and the converse part completes the proof. ~~~~~~~~~~~~~~~~~~~~~~~~~~~~~~~~~~~~~~~~~~~~~~~~~~~~~~~~~~ {\raggedleft{$\blacksquare$}}

\section{Proof of Theorem 4}
Fix the power allocation policy  $P^*$. Under $P^*(h)$, the achievability of  $\sup_{p_x: E[b(X)] \le P^*(h)} I(X;W)$, whenever $H_k=h$, is proved using  the techniques provided in Theorem 2  for the non-fading case.  Using this along with finding the expectation w.r.t. the optimum power allocation scheme completes the achievability proof.

The converse follows via Fano's inequality. ~~~~~~~~~~~~~~~~~~~~~~~~~~~~~~~~~~~~~~~~~~~~~~~~~~~~~~~~~~~~~~~~~~~~~~~~~~~~~~~{\raggedleft{$\blacksquare$}}

%\index{theorems}
%\scriptsize\parskip=10pt
\bibliographystyle{IEEEtran}

\bibliography{mybibfilefade_modi1}

%\bibliography{IEEEabrv,mybibfile}
\end{document}